\documentclass[11pt,twoside]{article}
\usepackage{graphicx}
\usepackage{txfonts}
\begin{document}
\begin{center}

\Large {\bf
Grain Alignment, Polarization and Magnetic Fields \\
}

\vspace{1pc}
\large

Alex Lazarian

\vspace{1pc}
\small
\it
Department of Astronomy, University of Wisconsin-Madison, 475 Charter 
St. Madison, WI 53705
\vspace{1pc}

\bf
Abstract
\rm
\vspace{1pc}
\small

\begin{minipage}[]{5in}
       Aligned non-spherical dust particles polarize starlight
       passing through the dust cloud. They also emit polarized
       far infrared and sub-mm radiation. 
       Substantial progress in understanding of grain alignment theory 
       makes the interpretation of the polarized radiation in terms
       of underlying magnetic fields much more reliable. I discuss
       a number of fundamental processes that affect grain alignment.
       In particular, I shall discuss how subtle effects related
       to nuclear spins of the atoms alter the dynamics
       of dust grains. I shall discuss how the theory explains the 
       existing observational data and demonstrate when the polarization
       can and cannot be interpreted in terms of the underlying magnetic        fields.   
 
\end{minipage}
\end{center}
\normalsize

\vspace{1pc}
\section{History of ideas}

Observations by \cite{Hall, Hiltner} revealed that interstellar
dust grains get aligned with respect to magnetic field. It did not take 
long time to realize that grains tend to be aligned with their long axes 
perpendicular to magnetic field. However, progress in theoretical 
understanding of the alignment
has been surprisingly slow in spite of the fact that great minds like
L. Spitzer and E. Purcell worked on the problem (see \cite{ST, P69, P75, P79, SM79}). 
Formulating the adequate grain alignment theory happened to be 
very tough and a lot of relevant physics had to be uncovered. 

Originally it was widely believed that interstellar grains can be well aligned
by a paramagnetic mechanism \cite{DG}. This mechanism based
on the direct interaction of rotating grains with the interstellar 
magnetic field required to have magnetic fields that are stronger than those
uncovered by other techniques.\footnote{As discussed for instance in 
\cite{Laz03} the very small grains are likely to be aligned by this mechanism and 
this can explain
the peculiarities of the UV part of the spectrum of the polarized 
radiation observed (see \cite{KM95}).
The efficiency of the Davis-Greenstein mechanism increases as the grain size 
decreases.}
Later, a pioneering work by \cite{P79} showed a way how to make grain 
alignment more
efficient. \cite{P79} noticed that grains rotating at high rates are not 
so susceptible
to the randomization induced by gaseous collisions and introduced 
several 
processes that
are bound to make grains very fast rotators. For decades this became a standard
explanation for grain alignment puzzle, although it could not explain several 
observational
facts, e.g. why observations indicate that small grains are less aligned 
than the large ones.

\section{Relevant Physics}

New physics of grain internal motion uncovered fairly recently 
explains inefficiency of alignment of small grains by Purcell's mechanism. 
\cite{LD99a} found that small grains flip frequently due to the
coupling of rotational and vibrational degrees of freedom of a grain. 
As the result
regular torques, e.g. torques due to ejection of H$_2$ molecules, 
get averaged out
and grains rotate at thermal velocities. The paramagnetic alignment of 
thermally
rotating grains as we mentioned earlier is inefficient 
(see \cite{RL99}).
Interestingly enough, \cite{LD99b} found that coupling 
of rotational
and vibrational degrees of freedom happens most efficiently through 
the so-called
nuclear relaxation that arises from nuclear spins of species within the grains.
This relaxation makes grains of size $\geq 10^{-5}$~cm rotate 
thermally,
which makes the Purcell mechanism inefficient for most of dust in diffuse
 interstellar medium.

\section{Radiative Torques}

 Introduced first in
\cite{Dog72} and \cite{DM76} the RT were
mostly forgotten till a more recent study \cite{DW96}, where their 
efficiency was demonstrated
using numerical simulations (see also \cite{DW97, WD03}, \cite{CL05b}).

The RT make use of interaction of radiation with a grain 
to spin the grain up. Unpolarized light can be presented as a superposition
of photons with left and right circular polarization. In general, the 
cross-sections of interaction of such photons with an irregular grain will
be different. As the result of preferential extinction of photons with
a particular polarization the grain experiences regular
torques and gets spun up.

The predictions of RT mechanism are roughly 
consistent
with the molecular cloud extinction and emission polarimetry 
\cite{LGM}
and the polarization spectrum measured \cite{Hild}. 
RT have been
demonstrated to be efficient in a laboratory setup \cite{Abbas}. 
Evidence in favor of RT alignment was found for the data obtained at the interface of the
dense and diffuse gas (\cite{Laz03} and Figure~1). 

While it was originally believed that RT cannot align grains
at optical depths larger than $A_v\approx 2$, a
recent work \cite{CL05b} demonstrated
that the efficiency of RT increases sharply with the grain size
and therefore bigger grains that exist within molecular clouds can be aligned
for $A_v$ more\footnote{The studies by one of us reveal that for fractal
molecular clouds the alignment can be present for cores with $A_v$ of
30. In addition, as large grains do not flip frequently the Purcell torques
and Purcell's alignment gets efficient as well.} than 10. 
Large grains may constitute an appreciable part
of the total mass of dust within a cloud, while still be marginal in terms
of light extinction. Therefore a non-detectable polarization in optical
and near infrared does not preclude substantial polarization to be present
in submillimeter. This makes submillimeter polarimetry the preferred tool
for studies magnetic fields and magnetic turbulence in molecular clouds.

\begin{figure}[htb]
\centerline{
\resizebox{\hsize}{!}{\includegraphics{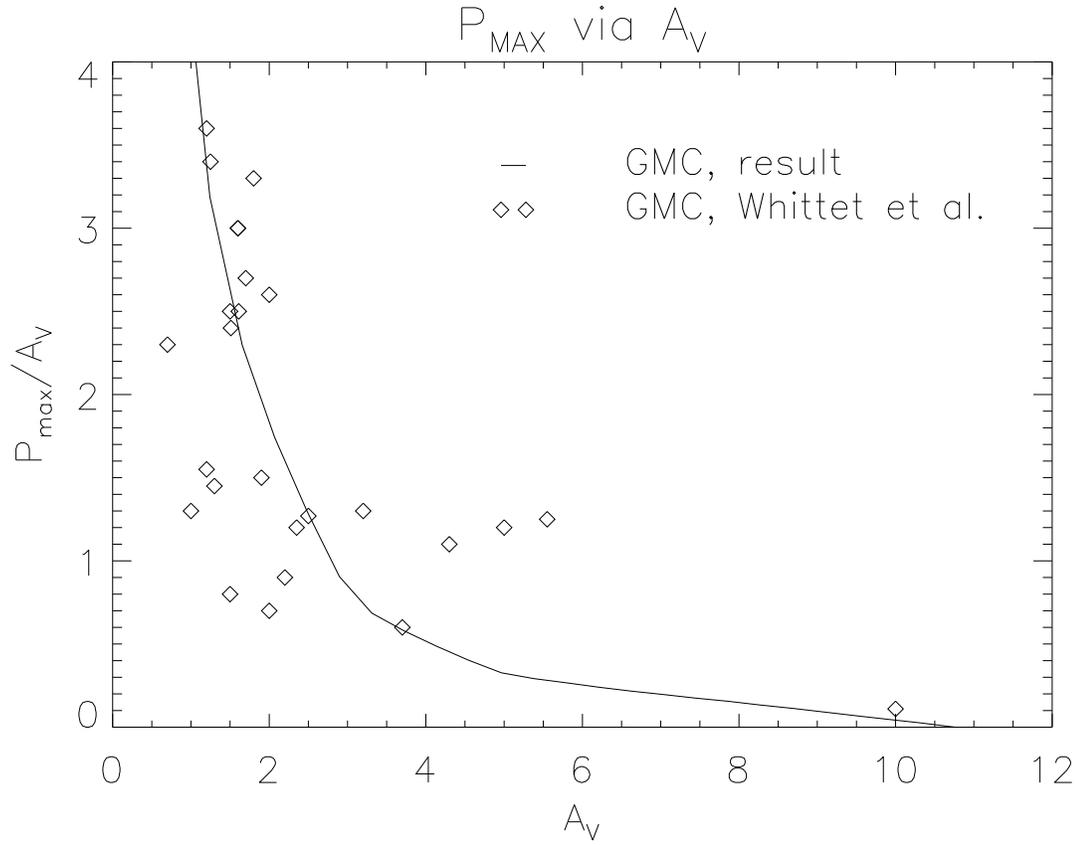}}
}
\caption
{$p_{max}/A_{V} $ as function of $A_{V}$ from our calculations
with radiative torques and the
observation data by Whittet et al. (2001). Observational error bars
are not shown. The flatterning of the observational results at
$A_v>4$ is likely to be the consequence of the cloud being inhomogeneous.
We use a homogeneous slab and the MRN distribution of dust with
$a_{max}=0.35 \mu m$. Work done with Mr.~Hoang Thiem.
}
\label{fig1}
\end{figure}


\section{Relation to Magnetic Field}

Most of the processes produce grain alignment in respect to magnetic
field, even if the alignment mechanism is of non-magnetic nature.
This is true due to the rapid precession of grains about magnetic field.
This precession stems from the substantial magnetic moment that grains get
due to their rotation \cite{DM76}. Indeed, a rotating
paramagnetic body gets a magnetic moment due to a Barnett effect. The 
corresponding period of grain precession $\tau_L$ 
is $\sim 10^5 B_{3}^{-1} a_{-5}^2$~s, where the 
the external magnetic field is normalized over its typical interstellar value
of $3\times 10^{-6}$~G and grain size is chosen to be $a=10^{-5}$~cm.
 This means that for turbulent motions on time
scales longer than $\tau_L$ grains orientation in respect to magnetic 
field lines does not change as the consequence of the adiabatic invariant
conservation. 

If the alignment happens on the time scales shorter than $\tau_L$ the dust
orientation may not reflect the magnetic field. For the RT
such a fast alignment will happen with longer grain axes perpendicular 
{\it to the direction of radiation}, while the fast mechanical alignment will
happen with longer axes parallel to the flow\footnote{The rule of thumb for
mechanical alignment is that it tends to minimize the grain cross section
for the grain-flow interaction, while for RT is that
the grain precession is minimized.}. The mechanical and  RT
alignment takes place on the time scale of approximately\footnote{The mechanical alignment happens faster due to the fact that the flows are supersonic. This is an important difference to be considered for transient alignment, but
has marginal consequences for the most of interstellar gas.} gaseous damping
time, which is for interstellar medium is 
$\sim 10^{11} T^{-1/2}_{100}n_{-20}^{-1} a_{-5}$~s, where 
typical temperatures and densities of cold interstellar medium, which
are respectively 100K and $20$~cm$^{-3}$ were used for the normalization.
Note, that magnetic alignment takes place over even longer time scales,
namely, $\sim 10^{13} B_{3} a^{2}_{-5}$. Therefore in most cases the magnetic
field indeed should act as the alignment axis.

It is worth noting that the turbulent fluctuations over time scales that are
shorter than $\tau_L$ {\it do not} suppress alignment. The rapidly precessing
grains preserve their orientation to the local direction of magnetic field
and undergo the alignment even when this local direction is changing its 
orientation in space. In this respect grain alignment is a local process
that can reflect local direction of magnetic field for magnetic ripples
larger than $\tau_L V_A$, where $V_A$ is Alfv\'en speed. Whatever is the process
of alignment, if it aligns grains over timescale larger than $\tau_L$ the
alignment is perpendicular or parallel to magnetic field. This allows to
study some aspects of magnetic turbulence without asking fundamental questions
about mechanisms of alignment.

\section{Summary}

Grain alignment is a subject that is very rich in terms of
physical processes. The advances in understanding
of grain alignment processes made the theory predictive and allowed
to explain the observational data available. This enables one to 
reliably interpret observed polarization in terms of the underlying magnetic
fields.

\subsection*{Acknowledgments}
This work is supported by the NSF grant AST 0243156 and the 
NSF Center for 
Magnetic Self-Organization in Laboratory and Astrophysical Plasmas.

\end{document}